\documentclass[]{spie}  
\usepackage[]{graphicx}

\title{CASTER - a concept for a Black Hole Finder Probe based on the use of new scintillator technologies} 

\author{Mark L. McConnell\supit{a}, Peter F. Bloser\supit{a}, Gary Case\supit{b,c}, Michael Cherry\supit{b}, James Cravens\supit{d},\\ T. Gregory Guzik\supit{b}, Kevin Hurley\supit{e}, R. Marc Kippen\supit{f}, John Macri\supit{a}, Richard S. Miller\supit{g}, \\William Paciesas\supit{g}, James M. Ryan\supit{a}, Bradley Schaefer\supit{b}, J. Gregory Stacy\supit{b,c},\\W. Thomas Vestrand\supit{f} and John P. Wefel\supit{b}
\skiplinehalf
\supit{a}Space Science Center, University of New Hampshire, Durham, NH  03824 ;\\
\supit{b}Department of Physics and Astronomy, Louisiana State University, Baton Rouge, LA  70803 ;\\
\supit{c}Department of Physics, Southern University, Baton Rouge, LA  70813 ;\\
\supit{d}Department of Space Science, Southwest Research Institute, San Antonio, TX  78228 ;\\
\supit{e}Space Sciences laboratory, University of California, Berkeley, CA 94720 ;\\
\supit{f}Los Alamos National Laboratory, Los Alamos, NM  87545 ;\\
\supit{g}Department of Physics, University of Alabama, Huntsville, AL  35899
}

\authorinfo{Send correspondence to: Mark.McConnell@unh.edu}

\begin{document} 
  \maketitle 

\begin{abstract}
The primary scientific mission of the Black Hole Finder Probe (BHFP), part of the NASA Beyond Einstein program, is to survey the local Universe for black holes over a wide range of mass and accretion rate. One approach to such a survey is a hard X-ray coded-aperture imaging mission operating in the 10--600 keV energy band, a spectral range that is considered to be especially useful in the detection of black hole sources. The development of new inorganic scintillator materials provides improved performance (for example, with regards to energy resolution and timing) that is well suited to the BHFP science requirements. Detection planes formed with these materials coupled with a new generation of readout devices represent a major advancement in the performance capabilities of scintillator-based gamma cameras. Here, we discuss the Coded Aperture Survey Telescope for Energetic Radiation (CASTER), a concept that represents a BHFP based on the use of the latest scintillator technology.
\end{abstract}



\section{INTRODUCTION}
\label{sect:intro}  

NASA's Beyond Einstein Program$^{1}$ defines a sequence of space missions for exploring of the Universe.  One aspect of this program is a series of three Einstein Probe missions that would complement the facility-class Einstein Great Observatories (LISA and Con-X).  Although the launch dates are highly uncertain at this time, it is hoped that the Probe missions will be launched in the 2012--2020 time frame.  One of the three Einstein Probe missions defined by Beyond Einstein roadmap is the Black Hole Finder Probe (BHFP).  The goal of the BHFP will be to carry out an all-sky census of accreting black holes:  from supermassive black holes in the nuclei of galaxies, to intermediate mass (about 100--1000 solar mass) holes produced by the very first stars, to stellar mass holes in our galaxy.  It is generally agreed that a hard X-ray coded mask imager covering the 10--600 keV energy band would be an effective tool for achieving this goal. 

One concept for the BHFP mission, known as EXIST (the Energetic X-ray Imaging Survey Telescope), has been under development for several years.$^{2-4}$  Here we offer an alternative concept, one that is similar to EXIST, but based on different detection technologies.  We will refer to our design concept as the Coded Aperture Survey Telescope for Energetic Radiation (CASTER).$^{5}$  CASTER is designed to employ several new experimental techniques using standard detector technologies, such as inorganic scintillators, wavelength-shifting fibers and photomultiplier tubes (PMTs), all of which have laboratory and space flight heritage. The development of a new inorganic scintillator material, lanthanum bromide (LaBr$_3$), provides improved performance that is well suited to the BHFP science requirements. Detection planes formed with LaBr$_3$ scintillator coupled with a new generation of readout devices may represent a major advancement in the performance capabilities of scintillator-based gamma cameras.  

The detector technology is an important driver for the mission design.  For example, limitations on detector thickness (i.e., detector efficiency) dictate the requirement on detector area, with all its implications for experiment size, weight, power, etc.  This in turn constrains the choices for launch vehicle and orbit. Scintillators represent a proven technology that is robust, reliable and simple to implement in large areas and in large volumes. With LaBr$_3$, we now have the prospect of scintillators with energy resolution and stopping power on par with room-temperature semiconductors (such as CZT), but with far less cost.   In addition, scintillator technology offers a practical means to extend the effective energy range beyond 511 keV, an important goal that may be difficult to achieve with the limited thicknesses of CZT. We therefore are exploring the implications, benefits and penalties, both practical and scientific, of using inorganic scintillators as the detector technology of a coded aperture imaging BHFP.

\section{MISSION REQUIREMENTS} 

The design goal of the Black Hole Finder Probe (BHFP) is to cover the full energy range from 10 keV up to 600 keV.  Many accreting black hole sources (especially stellar-mass black holes and Seyfert galaxies) have spectra that peak in this energy band.  This characteristic makes this particular energy range ideal for a BHFP mission.   The upper part of this energy range is of interest, in part, because it is sensitive to processes involving electron-positron annihilations.  Therefore, we place the upper limit of the energy range above the energy range of the 511 keV line. 

The BHFP sensitivity goal is defined for the 20-100 keV energy range so as to match the sensitivity of the ROSAT all-sky survey (RASS), $\sim0.05$ mCrab.  Carried out during the first six months of the ROSAT mission, the RASS covered the full sky in the soft X-ray range of 0.1--2.4 keV.  A total of 145,060 sources were detected by the RASS II analysis, 18,811 of which survived a screening process to make it into the ROSAT Bright Source Catalog (RBSC).$^{6}$

Although several wide-field instruments have been sensitive within the BHFP energy range, there has been only one instance of a systematic all-sky survey at these energies --- the 13--80 keV all-sky survey performed by HEAO-A4.  The HEAO-A4 survey recorded 72 sources at a flux sensitivity of 14 mCrab.$^{7}$  Within the next few years, the Swift hard X-ray all-sky survey$^{8}$ is expected to reach a sensitivity level of $\sim2$ mCrab, providing the first all-sky survey since HEAO A-4.

Several other instruments operating in the BHFP energy range have provided only partial sky coverage.   The SIGMA telescope on the GRANAT Observatory covered about one-fourth of the sky at energies between 35 keV and 1.3 MeV.  The flux sensitivity was better than 100 mCrab over the covered region and was as good as $\sim10$ mCrab in the galactic center region.$^{9}$ The first IBIS/ISGRI catalog compiled from a galactic plane survey$^{10}$ yielded $\sim$25 gamma-ray sources in the 25--160 keV energy band, at a flux sensitivity of $\sim3$ mCrab.

The angular resolution requirement is defined by the need to avoid source confusion.  With the sensitivity limit defined above, it is expected that up to $\sim$30,000 AGN will be detected by the BHFP (mostly in the lower part of the BHFP energy range).  In order to avoid source confusion at these levels, an angular resolution of 5--10$^{\prime}$ will be required.  Because we expect fewer sources to be detected at the higher energies, the resolution requirement will be more relaxed in the upper part of the BHFP energy range.

\section{IMPLEMENTATION} 

Although it may be possible to effectively cover the full energy range with a single instrument, the CASTER concept study is currently considering two separate imager concepts, one that will ensure adequate coverage at the highest energies and one that will ensure adequate coverage at the lowest energies.  Both are coded aperture imaging telescopes and both rely on the use of the latest in photon detection technologies.

\subsection{Coded Aperture Imaging}

Although other means of imaging high-energy radiation are possible, such as rotation modulation collimators or Fresnel zone plate imaging, only traditional coded aperture imaging$^{11,12}$ offers the wide FoV that is needed for the full-sky survey of the BHFP. The coded mask must be thick enough to provide a high contrast shadow on the detector plane. The measured shadow is used to construct the sky image, so the ability to measure the photon distribution on the detection plane will, in part, determine the quality of the resulting image.  A large area photon detection plane simultaneously measures both the position of the photon interaction (in three dimensions to avoid parallax effects) and the energy lost by the photon as a result of that interaction. The most notable satellite applications of this technique are GRANAT/SIGMA$^{13}$, BeppoSAX$^{14,15}$, INTEGRAL$^{16,17}$ and SWIFT$^{18,19}$ all of which follow in the pioneering footsteps of balloon-borne instruments such as DGT$^{20}$ and GRIP.$^{21,22}$

Several interrelated parameters must be considered in designing a coded aperture imaging system.  The angular resolution corresponds to the angular size of a mask element as seen from the detection plane, and so is dictated by the mask element size and the separation of the mask from the detector.  The mask-detector separation is constrained by the size of the launch vehicle fairing and the FoV requirements.  The detection plane must be able to resolve the individual mask elements in the projected pattern.  The exact ratio between detector spatial resolution and mask element size has an important effect on the S/N in the reproduced image.$^{23,24}$ For a fixed mask element size, improved spatial resolution results in an improvement in the imaging S/N.  It has been shown$^{24}$ that , to optimize the S/N using discrete detector elements, the mask element size (in both x and y) should be $\sim 1.5$ times larger than the detector spatial resolution.  
 
The size of the fully-coded FoV is determined by the geometric area of the detector and the mask-detector separation. Several factors will, however, limit the FoV. The most important of these are the mask and detector element geometries. The mask thickness must be sufficient to attenuate photons in the desired energy range. At the same time, the thickness of the mask must be limited to maintain uniformity of mask transmission for off-axis sources and avoid the resulting vignetting effects. Another important consideration for a large FoV instrument is the lateral distance traveled by a photon in the detector material before it interacts. This will depend not only on the angle of incidence of the photon, but also on its energy (i.e., its mean free path). The concern is that the photon may register in a detector element other than the element that it first encountered. This Òpixel crosstalkÓ effect can seriously degrade the imaging performance.  It can be minimized at a given energy by limiting the FoV (the photon incidence angle) or the detector thickness$^{25}$ or by measuring the depth of the interaction site within the detection plane.   Depth information can be used in constructing the image from different layers within the detector.  The ability to determine the depth of photon interaction is therefore a key requirement.

We have developed and tested the procedures for producing large coded aperture tungsten masks using relatively straightforward photolithographic etching techniques.$^{25}$ A mask pattern is produced which provides for the Òedge allowancesÓ required to produce the proper shaped holes.  The mask pattern is transferred from CAD to a silver-based mylar acetate film using a high resolution (16,000 dpi) laser printer.  Photoresist is rolled onto a tungsten sheet, and two acetate masks are applied to the front and back of the tungsten using double-sided tape.  The two masks are aligned using markers on the overlapping edges.  The tungsten-acetate sandwich is then illuminated with UV light which photoexcites the resist through the clear holes in the mask.  The acetate film is removed, the resist is developed in soda ash to harden the unexposed photoresist, and the excited polymer photoresist is washed away.  The piece is then etched in an acid ÒsoupÓ at a predetermined rate and temperature, rinsed with water, and stripped of the photoresist.  An X-ray image of a prototype mask, taken with 40 - 50 keV X-rays from a dental X-ray machine using a CsI microfiber array and CCD camera is shown in Fig. 1.

	Energetic photons and cosmic ray particles striking the tungsten will produce fluorescence photons at an energy just below that of the tungsten K-edge.  In order to attenuate the tungsten fluorescence photons, a graded mask must be used.   The choice of grading material will be dictated by the physical properties, ease of handling, cost of the material, and the imaging properties.  Tin and silver foils, for example, are both potentially useful.  A silver or tin mask layer can be etched, aligned (using the same procedure as for the acetate masks), and glued to the back of the tungsten in a graded sandwich to produce the final mask.  A thin layer of plastic scintillator may be placed below the mask for mechanical support and to provide a means of rejecting charged particles.

\begin{figure}
	\begin{minipage}[b]{.46\linewidth}
	\centering
	\includegraphics[width=2.75in]{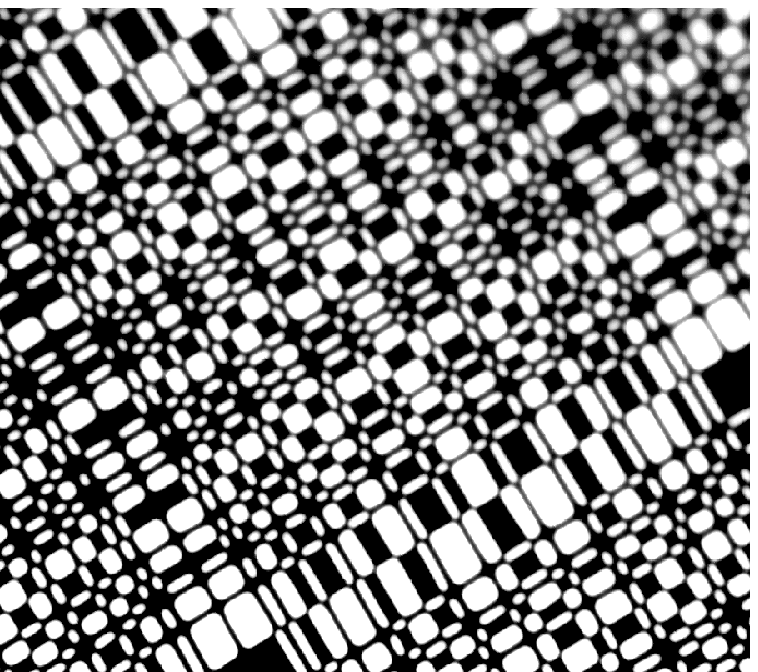}
	\caption{X-ray image of an etched tungsten mask obtained with a 1 mm thick CsI microfiber array and a CCD camera. The mask is a 71 x 73 URA pattern with 0.5 mm thickness and 0.5 mm minimum pixel dimensions.}
	\label{fig_sim}
	\end{minipage}\hfill
	\begin{minipage}[b]{.46\linewidth}
	\centering
	\includegraphics[width=3.00in]{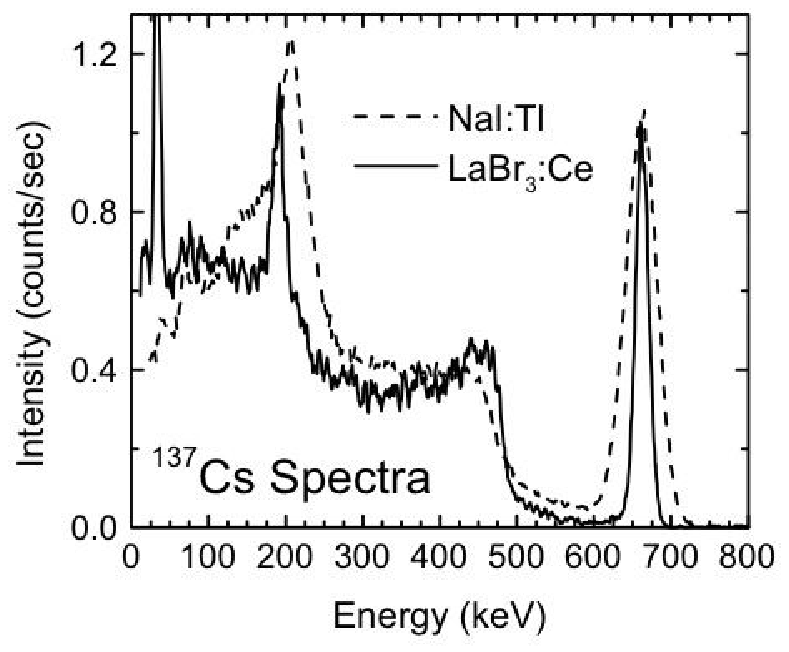}
	\caption{Energy spectrum of a $^{137}$Cs source obtained with a 1 cm$^3$ LaBr$_3$ detector.  The FWHM at 662 keV is 2.6\%.  The comparison NaI(Tl) spectrum has an energy resolution of 6.7\% (FWHM).}
	\label{fig_sim}
	\end{minipage}
\end{figure}

\subsection{Lanthanum Halide Scintillators}

Lanthanum halide scintillators (both Ce-doped Lanthanum Bromide, LaBr$_3$, and Ce-doped Lanthanum Chloride, LaCl$_3$) have become attractive alternatives for applications utilizing traditional scintillator methods.$^{26-31}$ Both materials offer the properties of high stopping efficiency, high light output, good linearity, significantly improved energy resolution, fast response, and (potentially) low cost. These materials, however, are still under development and there remain some technical issues to be addressed.  These include the ability to fabricate large crystal volumes, the intrinsic background (especially at energies above $\sim1$ MeV), and questions related to ruggedness, radiation hardness and activation.  In general, both LaBr$_3$ and LaCl$_3$ are attractive possibilities for CASTER.  We are currently baselining the use of LaBr$_3$, since it provides somewhat better performance characteristics, although LaCl$_3$, which is further along in its development, would be a suitable alternative.

\subsubsection{Scintillation Properties}

Both LaBr$_3$ and LaCl$_3$ offer significantly better scintillation light output than traditional NaI(Tl) scintillator.  Whereas the light output of NaI(Tl) is 38 photons/keV, LaBr$_3$ and LaCl$_3$ offer 63 and 49 photons/keV, respectively.$^{27,30}$  This level of light output is among the highest for inorganic scintillators.$^{32}$ The emission spectra of LaBr$_3$ and LaCl$_3$ (which peak at 380 and 350 nm, respectively) are well matched to the peak quantum efficiency (25\%) of borosilicate glass window PMTs with bialkali photocathodes.  The spectral match to such tubes is even better than that of NaI(Tl).  These emission spectra are also a good match to Saint Gobain waveshifting fibers BCF 91A, BCF 99-90 and BCF 99-33A, which have absorption spectra with peak wavelengths of 420, 345 and 375 nm, respectively.

The proportionality of light yield as a function of energy is another important property of any scintillator material.  All other factors equal, a more proportional scintillator will have a better energy resolution.$^{33}$ This is especially true at energies where one or more Compton interactions occur before the energy is fully absorbed.  Over the energy range from 60 to 1275 keV, the non-proportionality in light yield is about 6\% for LaBr$_3$ as compared to ~20\% for NaI(Tl) and CsI(Tl).$^{30}$  Results for LaCl$_3$ are similar to, but not quite as good as those for LaBr$_3$.

With their higher light output and better proportionality, both LaBr$_3$ and LaCl$_3$ exhibit significantly improved energy resolution.   Fig. 2 shows the spectrum of a $^{137}$Cs source obtained with a 1 cm$^3$ LaBr$_3$ detector at room temperature.$^{34}$ The energy resolution at 662 keV is 2.6\% FWHM. This result is comparable to the quoted energy resolution (3\% @ 662 keV) for off-the-shelf spectroscopy grade CZT (http://www.evproducts.com/) and is also comparable to the spectral resolution of the Swift CZT array.$^{18}$  However, at energies below $\sim100$ keV, the energy resolution of these materials appears to be comparable to NaI(Tl). The energy resolution advantage of these materials appears to be restricted to the upper part of the BHFP energy range.

The fluorescent decay times of both LaBr$_3$ and LaCl$_3$ are quite fast ($\leq30$ ns),$^{34}$ much faster than that of CsI (600--1000 ns), NaI(Tl) (230 ns) or BGO (300 ns).  This assures superior performance in high count-rate situations, where the smaller dead-time per event becomes a distinct advantage.  This will also be an advantage in efforts to reduce background via an active anticoincidence shield.  The coincidence timing resolution of LaBr$_3$ crystals has been measured with a FWHM of 260 ps,$^{34}$ a result which confirms that LaBr$_3$ is well suited for applications requiring fast response and good timing resolution.

\begin{figure}
	\begin{minipage}[t]{.46\linewidth}
	\centering
	\includegraphics[width=3.00in]{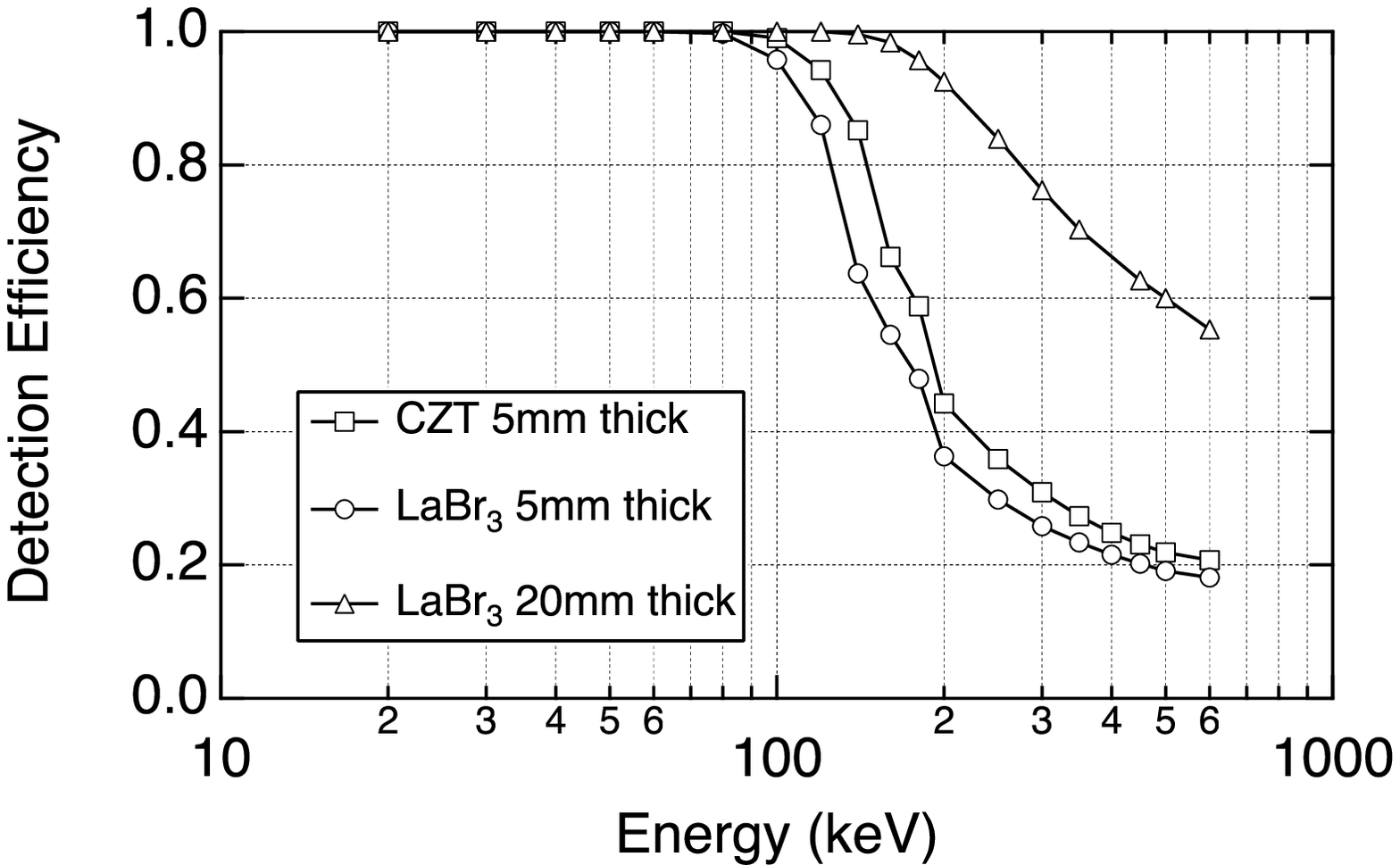}
	\caption{The total detection efficiency as derived from simulations of a cylindrical LaBr$_3$ scintillator of 1 cm radius and various thicknesses. }
	\label{fig_sim}
	\end{minipage}\hfill
	\begin{minipage}[t]{.46\linewidth}
	\centering
	\includegraphics[width=3.00in]{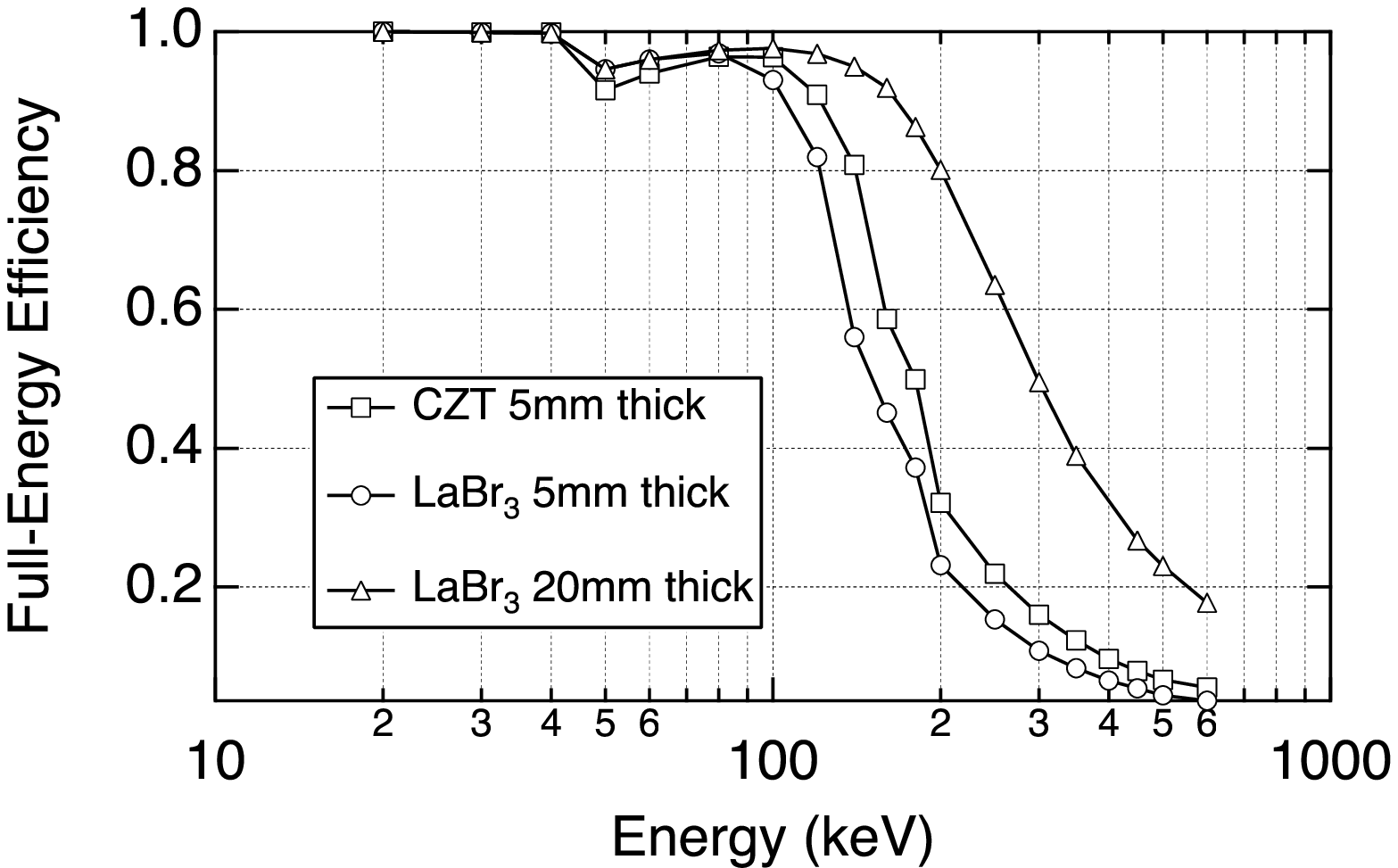}
	\caption{The full energy peak detection efficiency as derived from simulations of a cylindrical LaBr$_3$ scintillator of 1 cm radius and various thicknesses.}
	\label{fig_sim}
	\end{minipage}
\end{figure}

\subsubsection{Outstanding Issues}

Several experienced detector manufacturers are currently growing both LaCl$_3$ and LaBr$_3$ crystals using proprietary processes.  Small volume crystals of LaCl$_3$ (up to 2-inch size) are now routinely produced by Saint-Gobain and available as a stock item (under the trade name of BrilLanCe$^{{\tt TM}}$ 350).  While LaCl$_3$ crystals have been produced by Saint Gobain in sizes up to 3-inch diameter, the crystal sizes of the denser LaBr$_3$ material (BrilLanCe$^{{\tt TM}}$ 380) are somewhat more limited in size.  LaCl$_3$ development efforts lead those of LaBr$_3$ because it is slightly easier to work with. However, no fundamental barriers have been identified that would prevent crystal growth and detector fabrication with material volumes as large as are presently possible for NaI(Tl).  It is expected that, as production volumes increase, the cost of either material may also become comparable to that of NaI(Tl).

Lanthanum Halide detectors have been known to exhibit intrinsic backgrounds.$^{35}$  This has led to some concern about their use in astrophysical applications.  However, recent efforts to reduce the level of background in LaCl$_3$, based on techniques developed for ultra-low background NaI,  have been successful at reducing the intrinsic background between 1.5 and 2.5 MeV from $\sim$1-10 Bq/cc down to less than 0.05 Bq/cc.$^{36}$  The intrinsic background in the CASTER energy range appears not to be a major issue, although further study of this issue is warranted.

At the present time, there is very little known about how LaBr$_3$ or LaCl$_3$ will respond to intense doses of radiation.  CASTER is one of many applications where radiation damage and activation will be important issues.  We are currently planning tests to expose small samples of both materials to radiation beams during the fall of 2005.  In addition, we are planning to fly small samples of LaBr$_3$, LaCl$_3$ and NaI(Tl) as a piggyback payload on a balloon flight of the ATIC experiment later this year from Antarctica.  These data will provide us with the first measurements of the high altitude background of these scintillators.

\subsection{High Energy Telescope (60-600 keV)}

The concept for the CASTER High Energy Telescope (HET) is that of a large area, wide field-of-view coded mask imager based on the use of LaBr$_3$.  The photon detection plane will require some reasonable thickness of scintillator material ($\sim2$ cm) to insure adequate sensitivity at energies up to 600 keV.  In order to cover a reasonable fraction of the sky once every orbit, we require a total FoV for the HET of about $60^{\circ} \times 120^{\circ}$.  Two HET modules, each with a FoV of $70^{\circ} \times 70^{\circ}$ will satisfy this requirement.

The ability to fabricate LaBr$_3$ in large volumes (i.e., thicknesses greater than 1 cm) will offer the opportunity to provide improved detection efficiency at higher energies. Figs. 3 and 4 show data derived from simulations of detectors having a circular area with 1 cm radius.  The plots compare the data for 5 mm thick CZT with both 5 mm and 20 mm thick LaBr$_3$.  A thicker LaBr$_3$ detector can provide better sensitivity than a 5 mm thick CZT-based system, especially at energies above $\sim100$ keV (assuming comparable background levels). Improved detection plane sensitivity may also reduce the requirement on mask thickness. 

The angular resolution requirement of $\sim10$ arcminutes dictates the requirements for both the size of the coded mask elements and, subsequently, the spatial resolution of the detection plane.  Assuming that the mask-detector separation is 1.0 m, the coded mask element size must be $\sim3$ mm in order to satisfy the angular resolution requirement.  In order to resolve the mask shadow on the surface of the detection plane, we then require a (lateral) detector spatial resolution of 1--2 mm.  In order to avoid imaging aberrations, the depth of each photon interaction must also be measured with a comparable spatial resolution (1--2 mm).

The detection plane of the HET will consist of an array of Anger camera modules designed to measure the photon interaction location, along with the photon energy loss.  Such imaging detector arrays (or gamma cameras), formed with multiple light sensors viewing a layer of scintillation material, are widely used for medical imaging and other applications.$^{37,38}$  Most commercially available gamma cameras employ layers of NaI(Tl) scintillator viewed by arrays of PMTs.  The interaction location is determined for each detected photon from the relative signals recorded for each PMT in the array.  The spatial resolution of a gamma camera depends on geometrical factors such as the scintillator thickness, its size and the number and density of the light sensors in the array.  Statistical and optical factors, such as scintillation yield, surface reflectivity, spectral match and photoelectron yield are also important.  New gamma cameras employing the combination of higher light yield scintillator material (LaBr$_3$) and a higher density of light sensors, (e.g., small element MAPMTs) will have significantly better spatial resolution capabilities over earlier versions.  With proper calibration and analysis of the multiple PMT signals, the achievable energy threshold and spectroscopic performance of these Anger cameras should be similar to that achieved using single PMT spectrometers.$^{21,39}$

In addition to providing the location in the x- and y-dimensions, the extent of the distribution of scintillation signals within the sensor plane provides a measure of the z-coordinate or depth of the interaction.  Light from scintillations nearer the sensor plane is shared among fewer sensors than light from scintillations farther from the sensor plane. Fig. 5 illustrates the z-coordinate measurement principle.  Anger cameras with a higher density of readout sensors (with diameter and/or pitch less than the thickness of the scintillator layer) will have improved ability to measure the z coordinate. Multi-hit events would be identified and the interaction site locations measured in those cases where spatial resolution is better than the mean free path of the scattered photons.  This advantage is important for imaging above $\sim250$ keV where the fraction of Compton-scattered photons is greater. This will also open up the possibility of polarization studies with CASTER.

\begin{figure}
\centering
\includegraphics[width=3.0in]{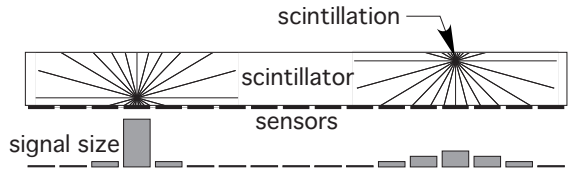}
\caption{Measurement of the photon interaction depth.}
\label{fig_sim}
\end{figure}

We have so far identified two attractive options for readout of the scintillation light signal.  One is a compact, low profile flat panel MAPMT (Hamamatsu H8500), designed specifically for large area arrays that is currently available in an $8 \times 8$ anode configuration with 5.6 mm anodes$^{40}$ (smaller anode versions of which are under development).  A second option is the Burle Planacon$^{\tt TM}$ tube,$^{41}$ based on potentially more rugged MCP technology, with a very similar geometry.  In either case, arrays of these devices have very little dead space ($\leq10$\%).  Other options, including  the use of discrete PMT arrays, may also be considered in the final HET design.

The use of a higher density of smaller sensor elements requires a larger number of processing channels that can be handled by the development of a suitable low power ASIC.  For example, the use of the Hamamatsu H8500 flat panel PMT (with an external size of 52 mm $\times$ 52 mm $\times$ 12.7 mm) would require roughly 28 $\times$ 28 or 784 such PMTs to cover the 1.5 m$^2$ area of the HET detection plane.  Assuming the use of a PMT with 64 anodes, this corresponds to more than 50,000 channels of data per HET module.

Image plane detectors formed with scintillators and MAPMTs will have significant power advantages over those formed with solid-state detectors.  The biasing of MAPMTs will require $< 1$ mW/cm$^2$ of detector area.  In addition, MAPMTs also provide high gain with little noise, thus reducing the preamplifier power required for each front-end electronics channel as compared to solid-state detectors.

\subsection{Low Energy Telescope (10-150 keV)}

The concept for the CASTER Low Energy Telescope (LET) is that of a large area, wide field-of-view coded mask imager based on the use of LaBr$_3$.  The LET will be optimized for the lower part of the CASTER energy range.  Because we anticipate a much higher density of sources at low energies, the angular resolution requirement for LET will be 5$^{\prime}$.   In order to cover the same total FoV as the HET modules ($60^{\circ} \times 120^{\circ}$), four LET modules, each with a FoV of $50^{\circ} \times 50^{\circ}$, will be required.

We are currently investigating an approach adapted from medical imaging applications$^{42,43}$ and analogous to the design of solid state strip detectors.  In this case, one layer of wavelength-shifting fibers is laid in the x-direction across the top of a scintillator and a second layer of fibers is laid in the y-direction across the bottom. The light emitted by the fibers is read out at one end of each fiber by a set of multi-anode photomultiplier tubes (MAPMTs).  The crossed fiber layers measure the x- and y- position using the center of gravity of the light in the two fiber arrays, and the depth by using the signal distribution across the fiber arrays. Only a small fraction of the light is absorbed, reemitted, and trapped in the fibers, however. Most of the light escapes the fibers. The energy measurement, therefore, is performed by a set of large area PMTs viewing the scintillator through the bottom fiber layer.  This approach offers the possibility to provide a significant reduction in the  power requirements by reducing the number of electronics channels as compared to a pixellated detector geometry.  Whereas the number of electronics channels for a gamma camera is on the order of $n_x n_y$, where $n_x$ and $n_y$ are the number of sensor elements in the $x$- and $y$-directions respectively, the number of electronics channels can be reduced to the order of $n_x + n_y$ by the use of this  Òcrossed fiberÓ readout approach.

In this configuration, the orthogonal geometry of the fiber layers permits the determination of both the $x$ and $y$ coordinates of the energy deposit.  When there is sufficient light output from the scintillator (e.g., for higher energy deposits), each interaction coordinate can be determined from the distribution of light within the fiber layer.  At lower energies, where the light output is reduced, the number of photoelectrons per fiber will become too small for the desired efficiency.  In order to overcome the limitations of this approach at low energies, segmented scintillator arrays may be used (as in Fig. 6).  The segmented nature of such arrays restricts the lateral spreading of the light within the scintillator.

Preliminary testing of this concept has recently been conducted at LSU using a 2.5 cm diameter $\times$ 2.5 cm thick LaBr$_3$ crystal.$^{44}$  The scintillation light is read out by a layer of 2 mm Saint Gobain BCF 99-33A fibers.  At the same time, the energy measurement is made by a single large area PMT that views the LaBr$_3$ through the fiber layer.  We expect some degradation of light yield (and, hence, energy resolution) as a result of measuring the light output through the fiber layer.  Tests show that, at 662 keV, the energy resolution of the test crystals decreases from 2.7\% for the case of the PMT directly viewing the LaBr$_3$ to 5.6\% in the case of the PMT viewing the crystal through the fiber layer.  These tests are continuing with the goal of more fully evaluating the energy resolution and spatial resolution that can be achieved using this approach.

\begin{figure}
\centering
\includegraphics[width=3.0in]{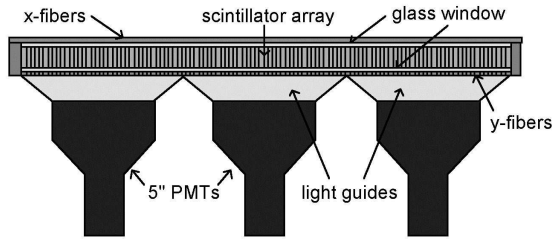}
\caption{An imager design consisting of a pixellated scintillator array read out by wavelength shifting fibers.  The large area PMTs are used for measuring total energy deposit.   A monolithic scintillator, rather than the pixellated array, could also be used in this configuration. }
\label{fig_sim}
\end{figure}

\subsection{Background Modeling}

	The sensitivity of CASTER to astronomical sources depends not only on the efficiency of the detection plane, but also on the level of background encountered on orbit. An accurate estimate of the in-flight background is therefore required in order to obtain meaningful sensitivity estimates. The primary sources of orbital background include$^{45-48}$ :

{\em Aperture photon flux} Ñ This includes all non-source photons that enter through the forward aperture.  The principal contributions come from the cosmic diffuse flux.  In a coded-mask telescope, half of these photons are stopped (or scattered) by the mask.

{\em Shield photon leakage} Ñ This includes photons that pass through the anticoincidence shield and are detected in the central detector without any shield signal.  As with the aperture flux, the principal contributions come from the Earth's atmosphere and the cosmic diffuse flux.

{\em Atmospheric neutrons} Ñ Neutron interactions within the experiment contribute to the background counting rate in at least two ways.  Thermal neutron capture and inelastic collisions often give rise to photons in our energy range.  In both cases, events in the 20--200 keV range can take place within the central detector.  The inelastic neutron scatters are particularly troublesome.  Typically, they excite a nucleus into a high-energy state or the continuum and the nucleus de-excites by emitting several photons.  This makes it important to have an active shield.  A passive shield must be very thick in order to stop all the secondary photons generated in the shield itself, whereas an active shield must only detect one secondary.

{\em Charged particles} Ñ Charged particles (both electrons and protons) passing through the central detector produce large signals.  These can be eliminated with active scintillators to detect (and reject) minimum ionizing particles.

{\em Effects of the coded mask} Ñ The presence of the coded mask can influence the background rate in several ways.  Background photons are generated in the mask by charged particles and high-energy photons.  Photons that might otherwise miss the detector may be scattered towards the detector by the mask.

We are currently working on the development of a  simulation model for CASTER based on the MGGPOD suite of simulation tools.$^{49}$  This should provide a reliable estimate of the CASTER in-flight background and will aid our efforts in the analysis of the CASTER design.

\subsection{Mission Architecture}

The CASTER mission concept closely parallels that of EXIST. The most significant change introduced by CASTER is in the imaging system.  In particular, CASTER envisions coded aperture detection planes based on a scintillator-based detector such as LaBr$_3$ rather than CZT.  This difference will have implications for the imaging system, background, mask fabrication, power requirements, and spacecraft weight.  All of these issues are being reconsidered in the current CASTER mission concept study.  The essential mission architecture, however, will remain largely unchanged.

The CASTER payload, as currently envisioned, will consist of six separate telescope modules.  There will be four low energy telescope (LET) modules and two high energy telescope (HET) modules.  The FoV of both the LET and HET arrays will be $60^{\circ} \times 120^{\circ}$.  The preliminary design specifications for the LET and HET modules are listed in Table 1.  The goal of locating bright sources to within 10$^{\prime\prime}$ implies that the CASTER aspect knowledge should be accurate to $\sim5^{\prime\prime}$.  The pointing stability need only be good to  a few arcminutes.  In its survey mode, lasting for perhaps one year, CASTER will be operated in a zenith-pointed mode.   Some rocking motion will be periodically induced to maximize the sky coverage.  In this way, CASTER will cover a significant fraction of the sky once each orbit.  The scanning mode offers the added benefit of dealing with non-uniformities in the detector background by effectively smoothing out the resulting image noise.  (INTEGRAL uses a dithering process to achieve the same result.)  Detector collimation may be required at lower energies to reduce the impact of the cosmic diffuse emission, which dominates the background at energies below $\sim$100 -- 200 keV. A combination of active and passive shielding will be used to minimize the effects of cosmic induced background and to provide Compton suppression. The preferred orbit will be a low-Earth equatorial orbit with a minimum two-year lifetime. An equatorial orbit will provide the lowest possible $\gamma$-ray background and therefore the highest sensitivity.  The HETE-2 mission has demonstrated the value of such a low-background orbit for high energy detectors.$^{50}$

To maximize the scientific return, the data will be returned on an event-by-event basis.  Each event will be time tagged with an accuracy of $\sim10 \mu$s.  Multi-hit information will be returned with each event to help construct the image and for polarization studies.  A dedicated burst response mode will be developed to provide prompt on-board processing of $\gamma$-ray burst data. On-board generation of appropriate sky maps will facilitate the rapid dissemination of burst locations to the astronomical community, as in the case of BATSE, HETE-2, and Swift.  Once the data are on the ground, rapid processing and release of the data to the public will enable rapid ground-based follow-ups.  A guest observer facility is envisaged to handle data dissemination and to provide support to guest investigators.

\begin{table}[h]
\caption{Preliminary CASTER telescope parameters.} 
\label{tab:fonts}
\begin{center}       
\begin{tabular}{|c|c|c|} 
\hline
\rule[-1ex]{0pt}{3.5ex}  & {\it HET}  &  {\it LET}\\
\hline
\rule[-1ex]{0pt}{3.5ex} Energy Range &   50-600 keV & 10-150 keV  \\
\hline
\rule[-1ex]{0pt}{3.5ex} No. of Modules & 2  &  4 \\
\hline
\rule[-1ex]{0pt}{3.5ex}  Detector Area$^{\dagger}$ & 1.5 m$^2$  &  1.5 m$^2$  \\
\hline
\rule[-1ex]{0pt}{3.5ex}  FoV$^{\dagger}$ & $\sim70^{\circ} \times 70^{\circ}$  & $\sim50^{\circ} \times 50^{\circ}$ \\
\hline
\rule[-1ex]{0pt}{3.5ex}  mask--detector sep & 100 cm & 150 cm  \\
\hline
\rule[-1ex]{0pt}{3.5ex}  angular resolution & $10^{\prime}$  & $5^{\prime}$ \\
\hline
\rule[-1ex]{0pt}{3.5ex}  mask element size & 3 mm  &  2 mm \\
\hline
\rule[-1ex]{0pt}{3.5ex}  mask element thickness &  7 mm &  1 mm  \\
\hline
\rule[-1ex]{0pt}{3.5ex}  total mask area$^{\dagger}$ & 3.0 m$^2$  & 3.0 m$^2$   \\
\hline
\rule[-1ex]{0pt}{3.5ex}  total mask weight$^{\dagger}$ & $\sim200$ kg   &  $\sim30$ kg  \\
\hline
\rule[-1ex]{0pt}{3.5ex}  detector spatial resolution  & $\sim 1-2$ mm  &  $\sim1$ mm\\
\hline
\rule[-1ex]{0pt}{3.5ex}  detector thickness  & 20 mm  &  10 mm\\
\hline
\rule[-1ex]{0pt}{3.5ex}  total detector weight$^{\dagger}$  & $\sim160$ kg  &  $\sim80$ kg \\
\hline
\end{tabular}
\par
$^{\dagger}$per detector module
\end{center}
\end{table}


\section{Conclusion}

Inorganic scintillator technology is well established in space-based applications.  LaBr$_3$ is a promising new scintillator material.  It has high stopping power, high light output, fast response, and shows good energy and timing resolution.  All studies to date indicate that these properties are maintained as the crystal volume is increased.  Undoubtedly the spatial resolution capabilities of gamma cameras made with LaBr$_3$ will be better than present day NaI or CsI-based instruments, but this remains to be demonstrated and the improvement measured.  The extent to which a higher density of scintillation light sensors improves the uniformity, 3-d spatial resolution, and multi-hit recognition capabilities of the detectors requires further study. In addition, a detailed demonstration of the crossed fiber performance, including position and energy resolution over the full range of incident photon energies and incidence directions, must be demonstrated. It will be important to further characterize detector capabilities, develop and validate LaBr$_3$ simulation models, and assess the impact of radiation exposures before defining optimum instrument configurations. 

\acknowledgements

This work has been supported in part by NASA grants NNG04GH78G and NNG05WC26G.

\end{document}